\documentclass{ws-ijmpd}
\usepackage[super,compress]{cite}

\usepackage{mathrsfs}
\usepackage{mathtools}
\usepackage{amsfonts}	
\usepackage{bm}			
\usepackage{hyperref}
\usepackage{subfigure}	
\usepackage{tensor}		
\usepackage{braket}		
\usepackage{tensor}		
\usepackage{xcolor}

\newcommand{\D}{\mathrm{d}}				
\newcommand{\hor}{\text{H}}				
\newcommand{\s}{\text{s}}				
\newcommand{\V}{\text{v}}				

\usepackage[nameinlink,capitalise]{cleveref}
\crefname{figure}{Figure}{Figures}

\begin{document}

\markboth{N.~M.~Santos, C.~L.~Benone, L.~C.~B.~Crispino, C.~A.~R.~Herdeiro and E. Radu}
{Stationary scalar and vector clouds around Kerr black holes}

%
\catchline{}{}{}{}{}
%

\title{STATIONARY SCALAR AND VECTOR CLOUDS\\AROUND KERR--NEWMAN BLACK HOLES}

%
%

\author{Nuno M. Santos,}
\address{Centro de Astrof\'{i}sica e Gravita\c{c}\~{a}o -- CENTRA and \\  Departamento de F\'{i}sica, Instituto Superior T\'{e}cnico -- IST \\ Universidade de Lisboa -- UL, Avenida Rovisco Pais 1, 1049, Lisboa, Portugal}
\author{Carlos A. R. Herdeiro}
\address{Centre for Research and Development  in Mathematics and Applications (CIDMA) and \\
Departamento de Matem\'atica da Universidade de Aveiro \\
Campus de Santiago, 3810-183 Aveiro, Portugal}

\maketitle

\begin{history}
\received{27 February 2020}
\end{history}

\begin{abstract}
Massive bosons in the vicinity of Kerr--Newman black holes can form pure bound states when their phase angular velocity fulills the synchronisation condition, \textit{i.e.} at the threshold of superradiance. The presence of these stationary clouds at the linear level is intimately linked to the existence of Kerr black holes with synchronised hair at the non-linear level. These configurations are very similar to the atomic orbitals of the electron in a hydrogen atom. They can be labeled by four quantum numbers: $n$, the number of nodes in the radial direction; $\ell$, the orbital angular momentum; $j$, the total angular momentum; and $m_j$, the azimuthal total angular momentum. These synchronised configurations are solely allowed for particular values of the black hole’s mass,
angular momentum and electric charge. Such quantization results in an existence surface in the three-dimensional parameter space of Kerr--Newman black holes. The phenomenology of stationary scalar clouds has been widely addressed over the last years. However, there is a gap in the literature concerning  their vector cousins. Following the separability of the Proca equation in Kerr(--Newman) spacetime, this work explores and compares scalar and vector stationary clouds around Kerr and Kerr--Newman black holes, extending previous research.
\end{abstract}


\ccode{PACS numbers:}


\section{Introduction}

Energy extraction from Kerr black holes was first devised in $1969$ by Penrose~\cite{Penrose:1969pc}, who conceived a \textit{gedankenexperiment} whereby a particle disintegrates within the ergo--region of a Kerr black hole into two other particles in such a way that the black hole loses energy. In general, the efficiency of the Penrose process is low: the extracted energy is at most about a fifth of the infalling energy for particles decaying close to the event horizon of extremal Kerr black holes~\cite{Wald:1974kya,Bardeen:1972fi}. More importantly, the minimum relative velocity between the two end-products of the decay must be greater than half the speed of light for energy to be extracted. The Penrose process is thus unlikely to occur and be relevant in conceivable astrophysical scenarios.

In 1971, Zel'Dovich showed that low--frequency electromagnetic waves scattered off a rotating conducting cylinder are amplified, later suggesting that, under particular circumstances, this enhancement occurs for any wave impinging on a rotating object~\cite{Zel'Dovich:1971,Zel'Dovich:1972}. Misner conjectured that Kerr black holes would not be an exception~\cite{Misner:1972}. This rather odd proposal opened the door to black--hole superradiance~\cite{Brito:2015oca}, which may be thought as the wave--analogue of the Penrose process.

For Kerr black holes, superradiance is triggered when the phase angular velocity $\omega$ of a boson state satisfies
\begin{equation}
\frac{\omega}{m_j}<\Omega_\text{H}\equiv\frac{a}{r_+^2+a^2}\ ,
\label{eq:1.1}
\end{equation}
where $m_j$ is the boson's azimuthal total angular momentum and $\Omega_\text{H}$ and $r_+=M+\sqrt{M^2-a^2}$ are, respectively, the black hole's horizon angular velocity and event horizon (Boyer-Lindquist) radial coordinate, written in terms of the black hole's ADM mass $M$ and total angular momentum $J=Ma$. When the bosons are massive, they remain trapped in the vicinity of the black hole -- as if they were enclosed by a reflective cavity. When Eq. \eqref{eq:1.1} is fulfilled, bosons extract energy from the black hole and, as a result, the trapped boson states grow exponentially with time, creating superradiant instabilities~\cite{Press:1972zz}. These arise even when the bosons' backreaction on the geometry is negligible -- a fairly good approximation for a plethora of astrophysical systems --, which means that superradiance is a linear phenomenon, although it persists at full non--linear level~\cite{East:2013mfa}.

From a dynamical viewpoint, energy extraction from the black hole stalls as soon as Eq. \eqref{eq:1.1} saturates, \textit{i.e.}
\begin{equation}
\frac{\omega}{m_j}=\Omega_\text{H} \ .
\label{eq:1.2}
\end{equation}
The endpoint is a classical boson condensate -- colloquially referred to as \textit{cloud} or \textit{hair} -- which is stationary with respect to the slowed-down black hole~\cite{Sanchis-Gual:2015lje,East:2017ovw,Herdeiro:2017phl}. These equilibrium configurations are solutions of Einstein's gravity minimally coupled to complex massive bosons, first unveiled for scalar bosons~\cite{Herdeiro:2014goa} and then extended to vector bosons~\cite{Herdeiro:2016tmi}. Kerr black holes with synchronised hair evade well-known uniqueness theorems~\cite{Herdeiro:2015waa} -- which state that asymptotically-flat stationary black holes in scalar-- or vector--(electro--)vacuum general relativity are necessarily Kerr(--Newman) black holes~\cite{Bekenstein:1971hc,Bekenstein:1972ky,Bekenstein:1972ny} -- and defy the \textit{no-hair} conjecture -- acoording to which the gravitational collapse in the presence of any type of matter-energy must give birth to a Kerr(--Newman) black hole\cite{Ruffini:1971bza,Cardoso:2016ryw} .

These hairy black holes reduce to synchronised bound states between Kerr black holes and (scalar or vector) bosons at the linear level. These states exist at the threshold of superradiance and are commonly known as stationary clouds. They are very similar to the atomic orbitals of the electron in a hydrogen atom in the sense that they are regular on and outside the event horizon, decay exponentially at spatial infinity and can be labeled by four quantum numbers: $n$, the number of nodes in the radial direction; $\ell$, the orbital angular momentum; $j$, the total angular momentum; and $m_j$, the projection of the total angular momentum
along the black hole's axis of symmetry.

These synchronised bound states were first found for massive scalar bosons around extremal ($a = M$) Kerr black holes~\cite{Hod:2012px} and later around rapidly-rotating black holes~\cite{Hod:2013zza}. While the phenomenology of stationary scalar clouds has been widely addressed in the literature over the last years~\cite{Hod:2014baa,Benone:2014ssa,Wang:2015fgp,Hod:2015goa,Siahaan:2015xna,Hod:2016lgi,Hod:2016yxg,Huang:2016qnk,Bernard:2016wqo,Sakalli:2016xoa,
Ferreira:2017cta,Richartz:2017qep,Huang:2017whw,Huang:2018qdl,Garcia:2018sjh,Delgado:2019prc,Kunz:2019bhm,Garcia:2019zla}, little is known about the physical properties of their vector cousins~\cite{Wang:2015fgp}. This discrepancy makes sense under the view that, as opposed to the Klein-Gordon equation~\cite{Teukolsky:1972my,Teukolsky:1973ha}, the decoupling and separation of the Proca equation in Kerr spacetime was solely achieved very recently via the Lunin--Frolov--Krtou\v{s}--Kubiz\v{n}\'{a}k (LFKK) ansatz~\cite{Frolov:2018ezx}. Following this breakthrough, which extends to the Kerr--NUT--(A)dS family of spacetimes, the properties of massive vector bosons started to be further explored in a number of spacetimes~\cite{Dolan:2018dqv,Cayuso:2019vyh,Baumann:2019eav,Siemonsen:2019ebd,Cayuso:2019ieu,Santos:2020pmh}, most notably the Kerr spacetime.  

The main goal of this paper is to apply the LFKK ansatz to characterize and compare stationary scalar and vector clouds around Kerr(--Newman) black holes, complementing some results presented in Refs.~\refcite{Benone:2014ssa} and~\refcite{Santos:2020pmh}.

The paper is organised as follows. \autoref{sec:2} reviews some key features of Kerr--Newman spacetime. \autoref{sec:3} introduces the Klein--Gordon and Proca equtions and the corresponding ans\"{a}tze for their separability and presents the separated equations. \autoref{sec:4} covers a comparative analysis of stationary scalar and vector clouds around Kerr and Kerr--Newman black holes.  An overview of the work is sketched in \autoref{sec:5}, together with some closing remarks.

\section{Kerr--Newman geometry\label{sec:2}}

The Kerr--Newman solution is the most general black--hole solution to the Einstein--Maxwell equations for an asymptotically--flat, stationary and axisymmetric spacetime with a connected event horizon. It describes a black hole with mass $M$, angular momentum $J$ and electric charge $Q$ (as measured from spatial infinity). A Kerr-Newman black hole is said to be sub-extremal if $a^2+Q^2<M^2$ and extremal if $a^2+Q^2=M^2$, where $a=J/M$ is the specific angular momentum. In Boyer--Lindquist coordinates $(t,r,\theta,\varphi)$, the solution reads
\begin{align}
\bm{g}&=\Sigma\left(-\frac{\Delta}{\Xi}\bm{\D}t^2+\frac{\bm{\D}r^2}{\Delta}+\bm{\D}\theta^2\right)+\frac{\Xi}{\Sigma}\sin^2\theta(\bm{\D}\varphi-\Omega\bm{\D}t)^2\ ,\nonumber\\
\bm{A}&-\frac{Qr}{\Sigma}(\bm{\D}t-a\sin^2\theta\bm{\D}\varphi)\ ,
\label{eq:2.1}
\end{align}
where
\begin{align}
&\Sigma=r^2+a^2\cos^2\theta\ ,
&&\Delta=r^2-2Mr+a^2+Q^2\ ,\nonumber\\
&\Xi=(r^2+a^2)^2-\Delta a^2\sin^2\theta\ ,
&&\Omega=\frac{(2Mr-Q^2)a}{\Xi}\ .
\label{eq:2.2}
\end{align}
The line element has a curvature singularity at $\Sigma=0$ and coordinate singularities at $\Delta=0$ when $a^2+Q^2\leq M^2$, which solves for $r=r_\pm\equiv M\pm\sqrt{M^2-a^2-Q^2}$. The hypersurface $r=r_+$ ($r=r_-$) is the outer (inner) horizon. 

Being stationary and axisymmetric, the Kerr-Newman spacetime does not depend explicitly on $t$ nor on $\varphi$. The two linearly independent Killing vectors associated with these two isometries are $\bm{\xi}=\bm{\partial}_t$ and $\bm{\eta}=\bm{\partial}_\varphi$, respectively. The Killing vector $\bm{\xi}$ is null on the hypersurface $r=r_\text{E}\equiv M+\sqrt{M^2-Q^2-a^2\cos^2\theta}$, known as stationary limit surface or ergosphere. This hypersurface is timelike except in the points in which $\bm{\eta}=0$, where it coincides with the outer horizon and becomes null. $\bm{\xi}$ is timelike outside the ergosphere and spacelike in the spacetime region between the outer horizon and the ergosphere ($r_+<r<r_\text{E}$). The points where $\bm{\eta}=0$ define the axis of symmetry.

The dragging potential $\Omega$ is constant on $r=r_+$, where it has the value
\begin{align}
\Omega_\hor\equiv\frac{a}{r_+^2+a^2}\ .
\label{eq:2.3}
\end{align}
$\Omega_\hor$ is thus the angular velocity of the outer horizon. The Killing vector $\bm{\chi}=\bm{\xi}+\Omega_\hor\bm{\eta}$ is null on the hypersurface $r=r_+$ and is timelike outside it. Observers moving along curves of constant $r$ and $\theta$ with angular velocity $\Omega_\hor$ follow the integral curves of $\bm{\chi}$ and thus rotate rigidly with the black hole.

The Kerr--Newman solution admits a principal tensor, \textit{i.e.} a non-degenerate closed conformal Killing--Yano $2$--form $\bm{h}$ which obeys the equations\footnote{The dot ($\cdot$) denotes contraction of two subsequent tensors with respect to their two
neighbor indices.}
\begin{align}
\bm{\nabla}\bm{h}=\bm{g}\wedge\bm{\xi}\ ,
\qquad
\bm{\xi}=\frac{1}{3}\bm{\nabla}\cdot\bm{h}\ .
\end{align}
This reads
\begin{align}
\bm{h}=r(\bm{\D}t-a\sin^2\theta\bm{\D}\varphi)\wedge\bm{\D}r-a\cos\theta\left[a\bm{\D}r-(r^2+a^2)\bm{\D}\varphi\right]\wedge\bm{\D}\cos\theta\ .
\end{align}
The Hodge dual of $\bm{h}$ is a Killing--Yano tensor $\bm{f}=\star\bm{h}$, whose square is the Killing tensor
\begin{align}
\bm{K}=-\bm{f}\cdot\bm{f}=\bm{h}\cdot\bm{h}-\frac{1}{2}\bm{g}h^2\ ,
\end{align}
which relates to the Killing vectors by $\bm{\eta}=\bm{K}\cdot\bm{\xi}$.

\section{Equations of motion\label{sec:3}}

The dynamics of massive scalar ($\Phi$) and vector ($\bm{A}$) bosons in curved spacetimes is ruled by similar equations:
\begin{align}
(\Box-\mu_\s^2)&\Phi=0\ ,\label{eq:3.1}\\
(\Box-\mu_\V^2)&\bm{A}=0\ ,\label{eq:3.2}
\end{align}
where $\Box\equiv\nabla_a\nabla^a$ is the D'Alembert operator and $\mu_s/\mu_\V$ stands for the mass of the scalar/vector boson. Equation \eqref{eq:3.1} is the Klein--Gordon equation, whereas Eq. \eqref{eq:3.2} is the Proca equation. The Proca equation is nothing but a set of four Klein--Gordon equations, supplemented by the Lorenz condition
\begin{align}
\bm{\nabla}\cdot\bm{A}=0\ ,
\label{eq:3.3}
\end{align}
which is automatically satisfied thanks to the non-vanishing mass $\mu_\V$.

It has long been known that the Klein-Gordon equation in Kerr--Newman spacetime allows a multiplicative separation of variables of the form~\cite{Carter:1968ks,Semiz:1991kh}
\begin{align}
\Phi(t,r,\theta,\phi)=e^{-i\omega t}R_\s(r)Q_\s(\theta,\phi)\ ,
\quad 
Q_\s(\theta,\phi)=S_\s(\theta)e^{+im_j\phi}\ ,
\label{eq:3.4}
\end{align}
where $\omega$ and $m_j$ are the eigenvalues of $i\bm{\xi}$ and $-i\bm{\eta}$, respectively. This ansatz reduces Eq. \eqref{eq:3.1} to two linear differential equations in the coordinates $r$ and $\theta$. These equations take the form
\begin{align}
\frac{\D}{\D r}\left[\Delta\frac{\D R_\s}{\D r}\right]+\left[\frac{K_r^2}{\Delta}-(\mu_\s^2r^2+a^2\omega^2-2m_ja\omega+\lambda_\s)\right]R_\s&=0\ ,\label{eq:3.5a}\\
\frac{1}{\sin\theta}\frac{\D}{\D\theta}\left[\sin\theta\frac{\D S_\s}{\D\theta}\right]-\left[\frac{m_j^2}{\sin^2\theta}-\nu^2\cos^2\theta-\lambda_\s\right]S_\s&=0\ ,\label{eq:3.5b}
\end{align}
where $K_r=(r^2+a^2)\omega-am_j$ and $\nu^2\equiv a^2(\omega^2-\mu_\s^2)$ is the degree of spheroidicity. Equations \eqref{eq:3.5a}--\eqref{eq:3.5b} are only coupled via the boson mass $\mu_s$, the Killing eigenvalues $\{\omega,m_j\}$, the black-hole parameters $\{M,a,Q\}$ and the separation constant $\lambda_\s$. When $\nu=0$ (\textit{i.e.} when the degree of spheroidicity vanishes), Eq. \eqref{eq:3.5b} reduces to the associated Legendre equation and the separation constant becomes $\lambda_s=j(j+1)$, $j\in\mathbb{N}_0$. The canonical solutions are the associated Legendre polynomials of degree $j$ and order $m$. The angular dependence of $\Phi$ is thus described by the scalar spherical harmonics of degree $j$ and order $m$ when either $a=0$ or $\omega^2=\mu_\s^2$. In general, however, it is given by scalar spheroidal harmonics. When $\nu\ll1$, the separation constant can be written as a series expansion around $\nu=0$,
\begin{align*}
\lambda_\s=\sum_{k=0}^{+\infty}f_\s^{(k)}\nu^{2k},
~
\text{with}
~
f_\s^{(0)}=\ell(\ell+1),
~
f_\s^{(1)}=h(\ell+1)-h(\ell)-1,\ldots,
\end{align*}
where $h(\ell)\equiv 2\ell(\ell^2-m^2)/(4\ell^2-1)$. Series expansions for large and real $\nu$ and for large and pure imaginary $\nu$ are also known~\cite{Berti:2005gp}. For generic degree of spheroidicity, \textsc{Mathematica} built-in function \textsc{SpheroidalEigenvalue}, for instance, retrieves high-precision results. 

The Proca equation, on the other hand, was believed not to separate in the Kerr--Newman spacetime. Until recently, the Hartle--Thorne formalism for slowly--rotating spacetimes~\cite{Hartle:1967he} was the only available (semi--)analytical technique to study massive vector bosons in Kerr spacetime\cite{Pani:2012vp,Pani:2012bp}. However, a new ansatz by Lunin for the separability of Maxwell's equations in the Myers-Perry-(A)dS family of spacetimes~\cite{Lunin:2017drx} was further developed by Frolov--Krtou\v{s}--Kubiz\v{n}\'{a}k \cite{Krtous:2018bvk,Frolov:2018pys}, who realised that the separability does extend to the Proca equation (and the Lorenz condition) in the Kerr--NUT--(A)dS family~\cite{Frolov:2018ezx}. The LFKK ansatz relies on the existence of hidden symmetries and allows Eq. \eqref{eq:3.1} to be separated into ordinary differential equations. This novel approach has already been applied to separate the torsion-modified Proca equation (known as Troca equation) in the Chong-Cveti\v{c}-L\"{u}-Pope spacetime of $D=5$ minimal gauged supergravity~\cite{Cayuso:2019vyh} and to study the superradiant instability of massive vector bosons in the Kerr--Newman and Kerr--Sen spacetimes~\cite{Cayuso:2019ieu}.

The LFKK ansatz for $\bm{A}$ takes the strikingly simple form
\begin{align}
\bm{A}=\bm{P}\cdot\bm{\nabla}Z\ ,
\label{eq:3.6}
\end{align}
where $\bm{P}$ is the \textit{polarisation tensor} and $Z$ is a complex scalar function. $\bm{P}$ is covariantly defined in terms of the metric $\bm{g}$ and the principal tensor $\bm{h}$ as
\begin{align}
\bm{P}\cdot\left(\bm{g}+\frac{i}{\lambda_\V}\bm{h}\right)=\bm{1}\ ,
\label{eq:3.7}
\end{align}
where $\lambda_\V$ is a complex constant and $\bm{1}$ is the $4$--dimensional identity matrix. Given the ansatz in Eq. \eqref{eq:3.6}, the Proca equation \textit{and} the Lorenz condition allow a multiplicative separation of variables for $Z$,
\begin{align}
&Z(t,r,\theta,\phi)=e^{-i\omega t}R_\V(r)Q_\V(\theta,\phi)\ ,\quad Q_\V(\theta,\phi)=S_\V(\theta)e^{+im_j\phi}\ ,
\label{eq:3.8}
\end{align}
where, as before, $\omega$ and $m_j$ are the eigenvalues of $i\bm{\xi}$ and $-i\bm{\eta}$, respectively. The separated equations in Kerr--Newman spacetime are\footnote{The explicit form of the polarisation tensor $\bm{P}$ in the Kerr(--Newman) spacetime can be found written in Boyer--Lindquist coordinates in Ref.~\refcite{Santos:2020pmh}.}
\begin{align}
q_r\frac{\D}{\D r}\left[\frac{\Delta}{q_r}\frac{\D R_\V}{\D r}\right]+\left[\frac{K_r^2}{\Delta}+\frac{2\lambda_\V^2-q_r}{q_r}\sigma\lambda_\V-q_r\mu_\V^2\right]R_\V=0\ ,\label{eq:3.9}\\
\frac{q_\theta}{\sin\theta}\frac{\D}{\D\theta}\left[\frac{\sin\theta}{q_\theta}\frac{\D S_\V}{\D \theta}\right]-\left[\frac{K_\theta^2}{\sin^2\theta}+\frac{2\lambda_\V^2-q_\theta}{q_\theta}\sigma\lambda_\V-q_\theta\mu_\V^2\right]S_\V=0\ ,\label{eq:3.10}
\end{align}
where
\begin{align}
&q_r=r^2+\lambda_\V^2\ ,\quad &&q_\theta=\lambda_\V^2-a^2\cos^2\theta\ ,\nonumber\\
&\sigma=a(m_j-a\omega)/\lambda_\V^2+\omega\ ,&&K_\theta=m_j-a\omega\sin^2\theta\ .
\label{eq:3.11}
\end{align}
Just like Eqs. \eqref{eq:3.5a}--\eqref{eq:3.5b}, Eqs. \eqref{eq:3.9}--\eqref{eq:3.10} are only coupled via the boson mass $\mu_\V$, the Killing eigenvalues $\{\omega,m_j\}$, the black-hole parameters $\{M,a,Q\}$ and the complex constant $\lambda_\V$. The latter may be loosely interpreted as a separation constant\footnote{In Ref.~\refcite{Frolov:2018ezx}, the authors first perform the separation of the Lorenz condition. This yields the separated equations, but with an additional constant, which is the actual separation constant. When separating the Proca equation, however, the new constant is fixed in terms of the boson mass $\mu_\V$ and the complex constant $\lambda_\V$. That is why $\lambda_\V$ can be referred to as a separation constant.}. Equation \eqref{eq:3.9} shares two singular points with Eq. \eqref{eq:3.5a}, $r=r_\pm$, and features additional poles at $r=\pm i\lambda_\V$. When $a=0$, Eq. \eqref{eq:3.9} reduces to the associated Legendre equation provided that $\lambda_{\V}^\text{E}(\lambda_{\V}^\text{E}-1)=j(j+1)$, which solves for $\lambda_{\V,-}^\text{E}=-j$ and $\lambda_{\V,+}^\text{E}=j+1$, where the superscript `E' refers to the electric--type states. Indeed, an asymptotic analysis of Eq. \eqref{eq:3.9} reveals that the angular dependence of the leading-order form of the spatial part of $\bm{A}$ is described by the electric--type `pure-orbital' vector spherical harmonics in flat space~\cite{Thorne:1980,Maggiore:2007}. More concretely, $\lambda_{\V,\mp}^\text{E}$ correspond to the $j=\ell\pm1$ electric--type states and can be written as a series expansion around $M\mu_\V=0$~\cite{Baumann:2019eav}, 
\begin{align}
\lambda^{\text{E}}_{\V,\pm}=\sum_{k=0}^{+\infty} f_{\V,\pm}^{(k)}(M\mu_\V)^k\ ,
\label{eq:3.12}
\end{align}
where
\begin{align*}
f_{\V,+}^{(0)}&=j+1\ ,~f_{\V,-}^{(0)}=-j\\
f_{\V,+}^{(1)}&=-\frac{m_ja}{j(j+1)M}\ ,~f_{\V,-}^{(1)}=\frac{m_ja}{jM}\ ,\ldots
\end{align*}
The magnetic--type states with $j=\ell=|m_j|$ can be recovered by taking the limits~\cite{Dolan:2018dqv}
\begin{align}
\lim_{M\mu_\V\rightarrow 0} \lambda^{\text{M}}_\V=0\ ,
\quad 
\lim_{M\mu_\V\rightarrow 0}\frac{\mu_\V a}{\lambda_\V^{\text{M}}}= m_j\pm 1\ , 
\label{eq:3.13}
\end{align}
where the superscript `M' refers to states with $j=|m_j|$. Unluckily, no series expansion of $\lambda_\V^{\text{M}}$ around $M\mu_\V=0$ is known. In the marginally-bound limit ($\omega^2=\mu_\V^2$), however, the separation constant takes the value
\begin{align}
\lim_{\omega^2\rightarrow\mu_\V^2}\lambda_\V^{\text{M}}=\frac{2a}{m_j+1-a\omega+\sqrt{(m_j+1-a\omega)^2+4a\omega}}\ ,
\label{eq:3.14}
\end{align}
which vanishes in the Schwarzschild limit. 

A potential caveat concerning the use of the LFKK ansatz is the fact that it might not capture all magnetic--type states. To the best of the authors' knowledge, only electric--type states and magnetic--type states with $j=|m_j|$ have so far been reported~\cite{Frolov:2018ezx,Dolan:2018dqv,Baumann:2019eav}. 

\section{Stationary scalar and vector clouds\label{sec:4}}

Quasi--bound states have frequencies whose real part is smaller than the boson mass $\mu$, Re$(\omega)<\mu$. Also, they behave as purely ingoing waves in the outer horizon's vicinity and decay exponentially at spatial infinity (as measured by a comoving observer), \textit{i.e.}
\begin{align}
\left. R\right|_{y\rightarrow-\infty}\sim e^{-i(\omega-m_j\Omega_\hor)y}\ ,
\quad
\left. R\right|_{y\rightarrow+\infty}\sim y^{-1}e^{-\sqrt{\mu^2-\omega^2}y}\ ,
\label{eq:4.1}
\end{align}
where the subscripts `$\s$' and `$\V$' were (and will hereafter be) omitted to avoid clutter and  $y$ is the tortoise coordinate, defined by
\begin{align}
y(r)=r+\frac{r_+^2}{r_+-r_-}\log(r-r_+)-\frac{r_-^2}{r_+-r_-}\log(r-r_-)\ .
\label{eq:4.2}
\end{align}
These states can be labelled by four `quantum' numbers: $n\in\mathbb{N}_0$, the number of nodes in the radial direction; $\ell$, the orbital angular momentum; $j$, the total angular momentum; and $m_j$, the projection of the total angular momentum along the black hole's axis of symmetry, which defines the number of nodes in the azimuthal direction. In general, only $n$ and $m_j$ are legitimate `quantum' numbers in the sense that they describe values of conserved quantities. Both orbital and total angular momenta are not conserved in rotating spacetimes. However, it is still convenient to use $\ell$ and $j$ to label scalar ($j=\ell$) and vector ($j=\ell-1,\ell,\ell+1$) states, always bearing in mind that they are only physically meaningful in Minkowski spacetime.   

When the (phase angular velocity of the) boson and the (horizon angular velocity of the) black hole synchronise, the oscillatory behavior close to the outer horizon vanishes and the resulting radial profiles become similar to those of the atomic orbitals of the electron in a hydrogen atom. In the following, synchronised scalar and vector states will be labelled with $\ket{n,j,m_j}$ and $\ket{n,\ell,j,m_j}$, respectively. 

These synchronised states are only supported by Kerr--Newman black holes in a particular domain of the 3--parameter space described by the dimensionless quantities $\{M\mu,a\mu,Q\mu\}$ or, equivalently\footnote{Equations \eqref{eq:3.5a} and \eqref{eq:3.9} can be written in terms of the black-hole parameters $\{r_+,a,Q\}$ using $\Delta(r_+)=0$.}, $\{r_+\mu,a\mu,Q\mu\}$ -- the latter is the gauge used in this work. A simple direct--integration shooting method~\cite{Pani:2013pma} suffices to scan the parameter space in search of synchronised (scalar and vector) states. To impose the desired behavior close to the outer horizon $R$ is written as a series expansion around $r=r_+$,
\begin{align}
\left.R\right|_{r\rightarrow r_+}\sim\sum_{k=0}^{+\infty}c_{(k)}(r-r_+)^k\ ,
\label{eq:4.3}
\end{align} 
where $c_{(0)}=1$ and the coefficients $\{c_{(k)}\}_{k>0}$ are obtained by solving either Eq. \eqref{eq:3.5a} or \eqref{eq:3.9} order by order. The coefficients depend on the boson mass $\mu$, the Killing eigenvalues $\{\omega,m_j\}$, the black-hole parameters $\{r_+,a,Q\}$ and the corresponding separation constant. Fixing $\{\ell,j,m_j\}$ and the black-hole parameters  $\{r_+\mu,Q\mu\}$, for instance, Eq. \eqref{eq:3.5a} or \eqref{eq:3.9} is then integrated from $r=r_+(1+\delta_+)$, with $\delta_+\ll 1$, to $r=r_{\infty}$, where $r_{\infty}$ stands for the numerical value of the radial coordinate at spatial infinity. 

Numerical solutions with the appropriate boundary conditions at spatial infinity are found via the shooting method. They only exist for discrete values of the specific angular momentum $a$, each corresponding to a different node number $n$. In other words, bound states between Kerr--Newman black holes and synchronised states are thus restricted to closed surfaces in the 3-parameter space spanned by $\{M\mu,a\mu,Q\mu\}$. Fixing $Q\mu$, for instance, these surfaces reduce to line segments in the 2-parameter space spanned by $\{M\mu,a\mu\}$ or, alternatively, $\{M\mu,\Omega_\hor/\mu\}$. These are commonly known as \textit{existence lines}. This paper's main goal is to determine and compare the existence lines of synchronised scalar and vector states around Kerr and Kerr--Newman black holes. The defining features of these lines will be outlined, without loss of generality, for the Kerr spacetime ($Q\mu=0$). This is also particularly convenient for the reader to compare the results presented herein with those already reported in the literature~\cite{Herdeiro:2014goa,Benone:2014ssa,Herdeiro:2016tmi,Santos:2020pmh}. The subtleties introduced by a non-vanishing electric charge will then be briefly addressed.

\subsection{Kerr black holes}

When the black hole's gravitational radius, $R_\text{G}=M$, is much smaller than the boson's reduced Compton wavelength, $\lambda_\text{C}=\mu^{-1}$, the frequency spectra of scalar and vector quasi--bound states in Kerr spacetime can be written in the form~\cite{Baumann:2018vus,Baumann:2019eav,Baumann:2019ztm}
\begin{align}
\omega^{\text{(s)}}_{\ket{n,j,m_j}}&=\mu\left(1-\frac{\alpha^2}{2\mathfrak{n}^2}-\frac{\alpha^4}{8\mathfrak{n}^4}+\frac{g(n,j,j)}{\mathfrak{n}^3}\alpha^4+\frac{h(j,j)}{\mathfrak{n}^3}\frac{m_j a}{M}\alpha^5+\ldots\right)\ ,\label{eq:4.3}\\
\omega^{\text{(v)}}_{\ket{n,\ell,j,m_j}}&=\mu\left(1-\frac{\alpha^2}{2\mathfrak{n}^2}-\frac{\alpha^4}{8\mathfrak{n}^4}+\frac{g(n,\ell,j)}{\mathfrak{n}^3}\alpha^4+\frac{h(\ell,j)}{\mathfrak{n}^3}\frac{m_j a}{M}\alpha^5+\ldots\right)\ ,
\label{eq:4.4}
\end{align}
where $\alpha=M\mu\ll1$ is the so--called gravitational fine--structure constant, $\mathfrak{n}\equiv n+\ell+1$ ($\mathfrak{n}\in\mathbb{N}$) may be referred to as principal quantum number and
\begin{align*}
&g(n,\ell,j)=-\frac{4(6\ell j+3\ell+3j+2)}{(\ell+j)(\ell+j+1)(\ell+j+2)}+\frac{2}{n+\ell+1}\ ,\\
&h(\ell,j)=\frac{16}{(\ell+j)(\ell+j+1)(\ell+j+2)}\ .
\end{align*}
Note that $\omega^{\text{(v)}}_{\ket{n,j,j,m_j}}=\omega^{\text{(s)}}_{\ket{n,j,m_j}}$, which suggests that the magnetic--type vector states are somehow equivalent to the scalar states with the same total angular momentum. However, it is worth pointing out that, as opposed to the frequencies of the electric--type states, computed analytically via matched asymptotic expansions, Eq. \eqref{eq:4.4} with $j=\ell$ is nothing but a conjecture. Nevertheless, all approximations are fairly accurate when $\alpha\lesssim 0.2$, even for near-extremal Kerr black holes~\cite{Baumann:2019ztm}.

The instability rates of the quasi--bound states are proportional to the factor $\text{sign}~w$, where $w\equiv(\omega-m_j\Omega_\text{H})$, \textit{i.e.} the states: grow exponentially with time when $\text{sign}~w=-1$, thus being unstable; decay exponentially with time when $\text{sign}~w=+1$, thus being stable; and are stationary (infinitely long--lived) when $w=0$. The unstable states are superradiant, while the stable states are non-superradiant. The stationary states, which are synchronised with the black hole, exist precisely at the threshold of superradiance. Their existence lines will be presented for fixed values of $Q\mu$ in the $(M\mu,\Omega_\hor/\mu)$-plane, in which the existence domain of Kerr(--Newman) black holes is shaded light gray. Note that the contour lines for which 
\begin{align}
\omega^{\text{(s)}}_{\ket{n,j,m_j}}&=m_j\Omega_\text{H}\ ,\\
\omega^{\text{(v)}}_{\ket{n,\ell,j,m_j}}&=m_j\Omega_\text{H}
\label{eq:4.5}
\end{align}
constitute an analytical approximation to the existence lines of scalar and vector states, respectively. A comparision between \textit{analytical} and \textit{numerical} existence lines of some vector states can be found in Ref.~\refcite{Santos:2020pmh}. Overall, the agreement is excellent, except when $j=m_j$ and $\ell<j$.

When $Q\mu=0$, the existence lines cover the entire range of the specific angular momentum $a$, with the endings matching the Schwarzschild ($a=0$) and extremal ($a=M$) limits. The former (latter) coincides with the minimum (maximum) allowed value for the gravitational fine--structure constant $\alpha=M\mu$.  

\begin{figure}[t]
  \centering
  \includegraphics[width=.7\linewidth]{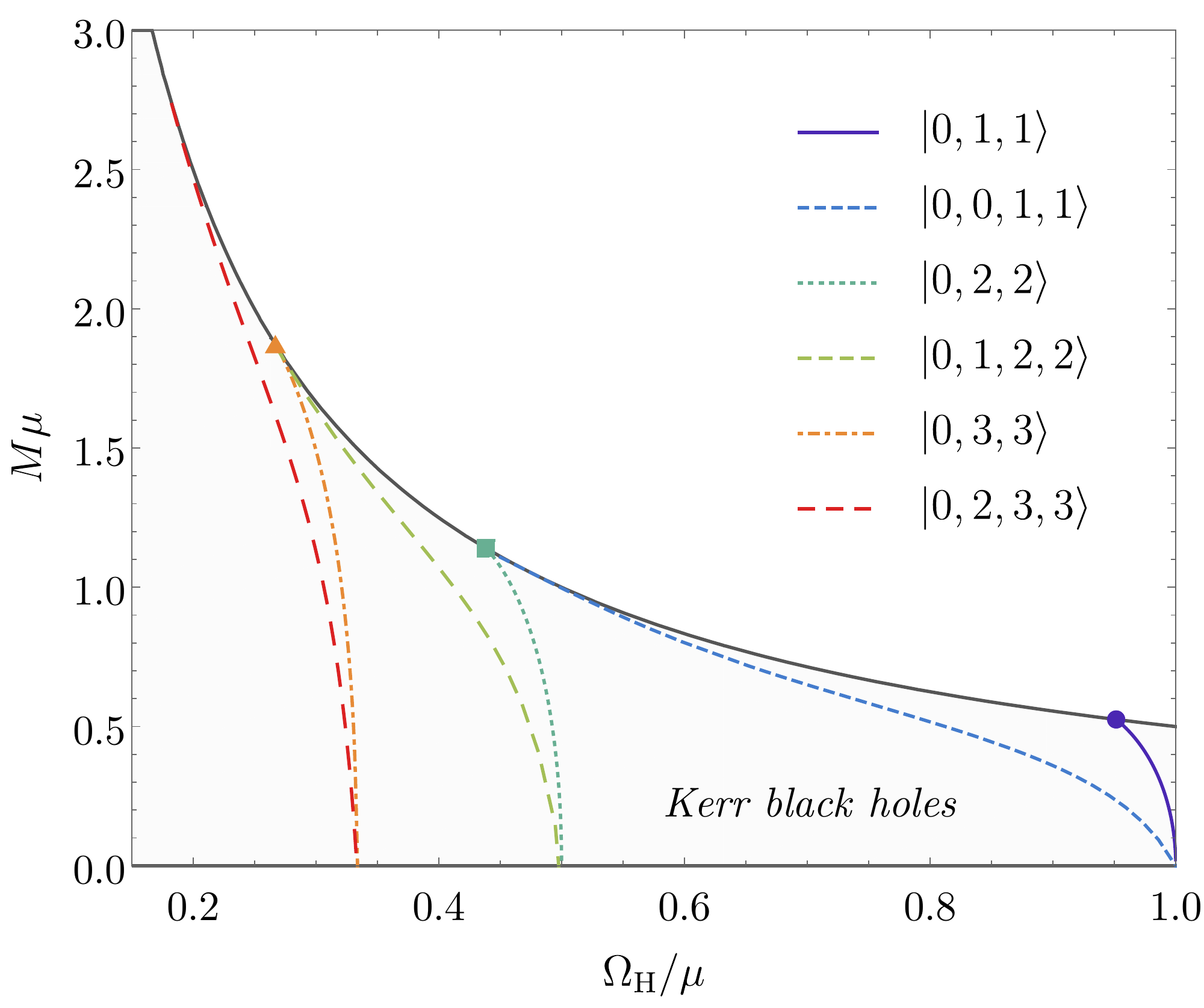}
  \caption{Existence lines of the first fundamental scalar and vector states with $j=m_j$ in the ($M\mu,\Omega_\text{H}/\mu$)--plane. The gray solid line refers to extremal ($a=M$) Kerr black holes. The markers pinpoint extreme ($a=M$) scalar states found analytically~\cite{Hod:2012px}. The vector states are less energetic than the corresponding scalar states, as they correspond to lower values of $\Omega_\hor$.}
  \label{fig:fig1}
\end{figure}

For a given $m_j$, the fundamental state does not possess any node in the radial direction and always has its total angular momentum completely aligned with the black hole's axis of symmetry ($j=m_j$). $\ket{0,m_j,m_j}$ thus represents fundamental scalar states. The fundamental vector states are those which cumulatively have the smallest possible orbital angular momentum, which corresponds to the electric--type states $\ket{0,m_j-1,m_j,m_j}$. The existence lines for the fundamental scalar and vector states with $m_j=1,2,3$ are shown in \autoref{fig:fig1}, where the markers pinpoint extreme ($a=M$) scalar states obtained by solving analytically Eq. \eqref{eq:3.5a} in terms of confluent hypergeometric functions~\cite{Hod:2012px}. These particular existence lines represent the threshold between Kerr black holes which are stable against all states with a given $m_j$ and the ones which are unstable against at least one such state. Fundamental vector states  always lie to the left with respect to the scalar state with the same $m_j$. The $\Omega_\hor$--interval of the vector states are greater than that of the corresponding scalar cousins -- e.g. it is approximately ten times greater for $\ket{0,0,1,1}$ than for $\ket{0,1,1}$. Put it differently, for a given boson mass $\mu$, Kerr black holes with sufficiently small horizon angular velocity may support vector, but not scalar states. These properties are a natural manifestation of the difference in strength of the superradiant instability, which is stronger for massive vector bosons~\cite{Press:1972zz}.  

\begin{figure}[h!]
  \centering
  \includegraphics[width=.70\linewidth]{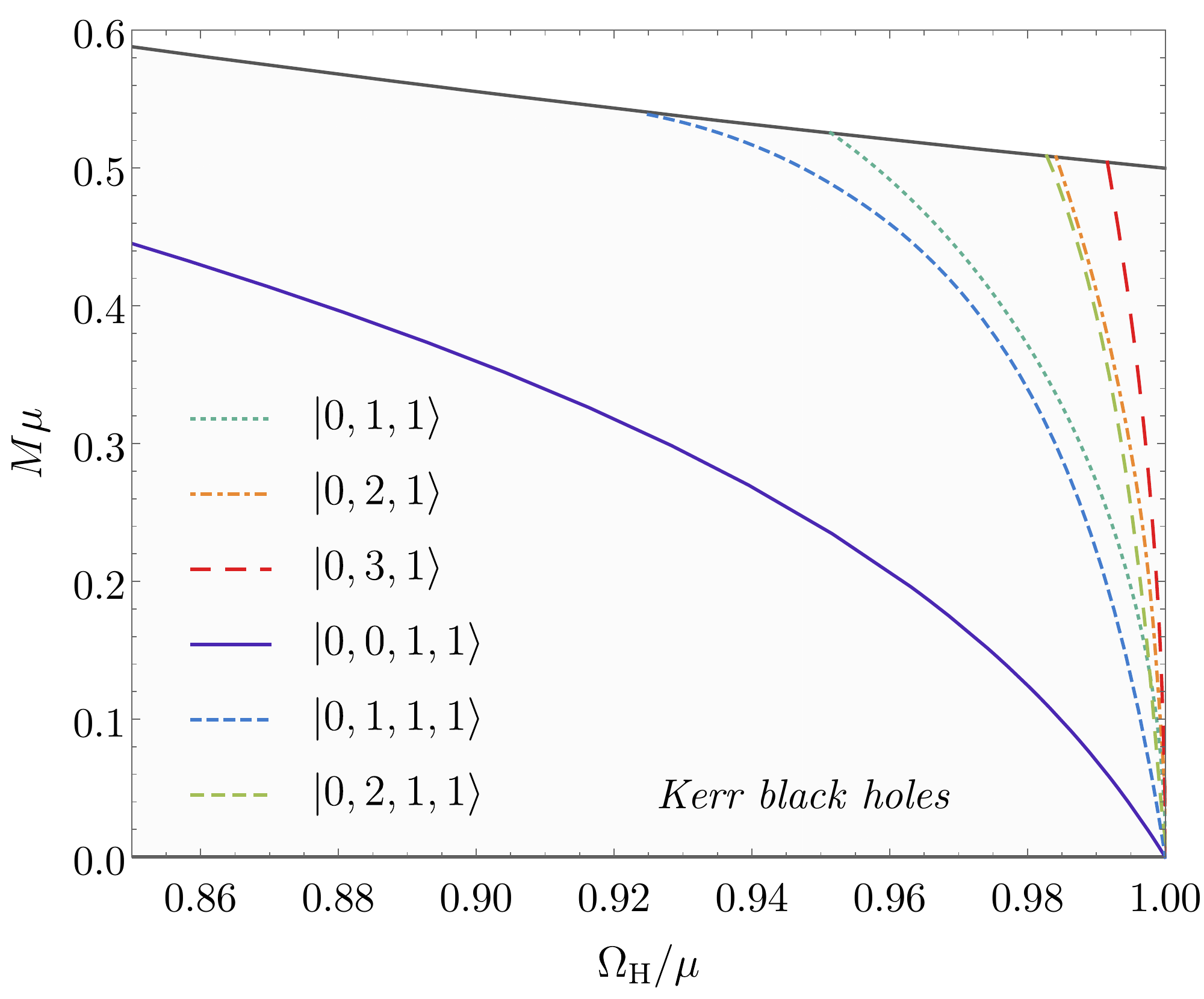}
  \caption{Existence lines of the scalar states $\ket{0,\ell,1}$, $\ell=1,2,3$ and vector states $\ket{0,\ell,1,1}$, $\ell=0,1,2$ in the ($M\mu,\Omega_\text{H}/\mu$)--plane. The gray solid line refers to extremal ($a=M$) Kerr black holes. The $\ell=0$ vector states are the least energetic, as they correspond to lower values of $\Omega_\text{H}$. The energy increases with $\ell$.}
  \label{fig:fig2}
\end{figure}

Excited states, on the other hand, must lie to the right with respect to the corresponding fundamental states in the $(M\mu,\Omega_\hor/\mu)$--plane. For example, fixing $\{n,j,m_j\}$, existence lines migrate towards greater and greater horizon angular velocities as $\ell$ increases. This behavior is illustrated in \autoref{fig:fig2}. The impact of the orbital angular momentum on the existence lines is particularly relevant for near--extremal Kerr black holes. In the Schwarzschild limit, the lines converge to $(M,\Omega_\hor)=(0,\mu)$, which amounts to saying that Schwarzschild black holes do not admit synchronised (scalar nor vector) bound states~\cite{Herdeiro:2014goa}. \autoref{fig:fig2} also shows the energy ordering of vector states with fixed $j$: the electric-type states $\ket{n,j+1,j,m_j}$ are more energetic than the magnetic-type states $\ket{n,j,j,m_j}$ and the latter more energetic than the electric-type states $\ket{n,j-1,j,m_j}$. This hierarchy matches the one found in the frequency spectrum of vector quasi--bound states~\cite{Baumann:2019eav}. 

\begin{figure}[h]
  \centering
  \includegraphics[width=.7\linewidth]{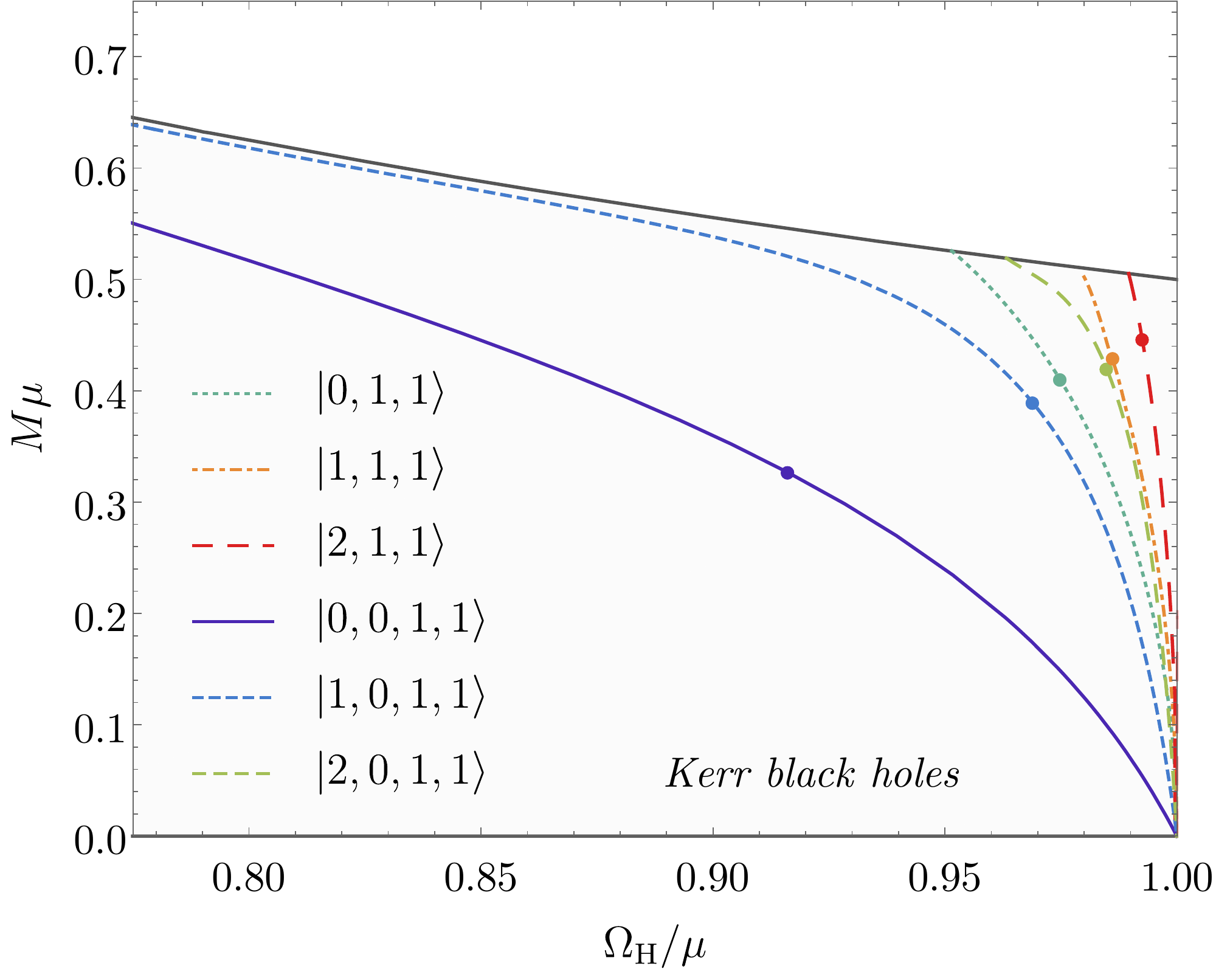}
  \caption{Existence lines of the scalar states $\ket{n,1,1}$ and vector states $\ket{n,0,1,1}$, $n=0,1,2$ in the ($M\mu,\Omega_\text{H}/\mu$)--plane. The gray solid line refers to extremal ($a=M$) Kerr black holes. The $n=0$ states are the least energetic, as they correspond to lower values of $\Omega_\text{H}$. The energy increases with $n$. The plot markers are states with $r_+\mu=0.5$, whose radial profiles are shown in~\autoref{fig:fig4}.}
  \label{fig:fig3}
\end{figure}

\begin{figure}[h]
  \centering
  \includegraphics[width=.7\linewidth]{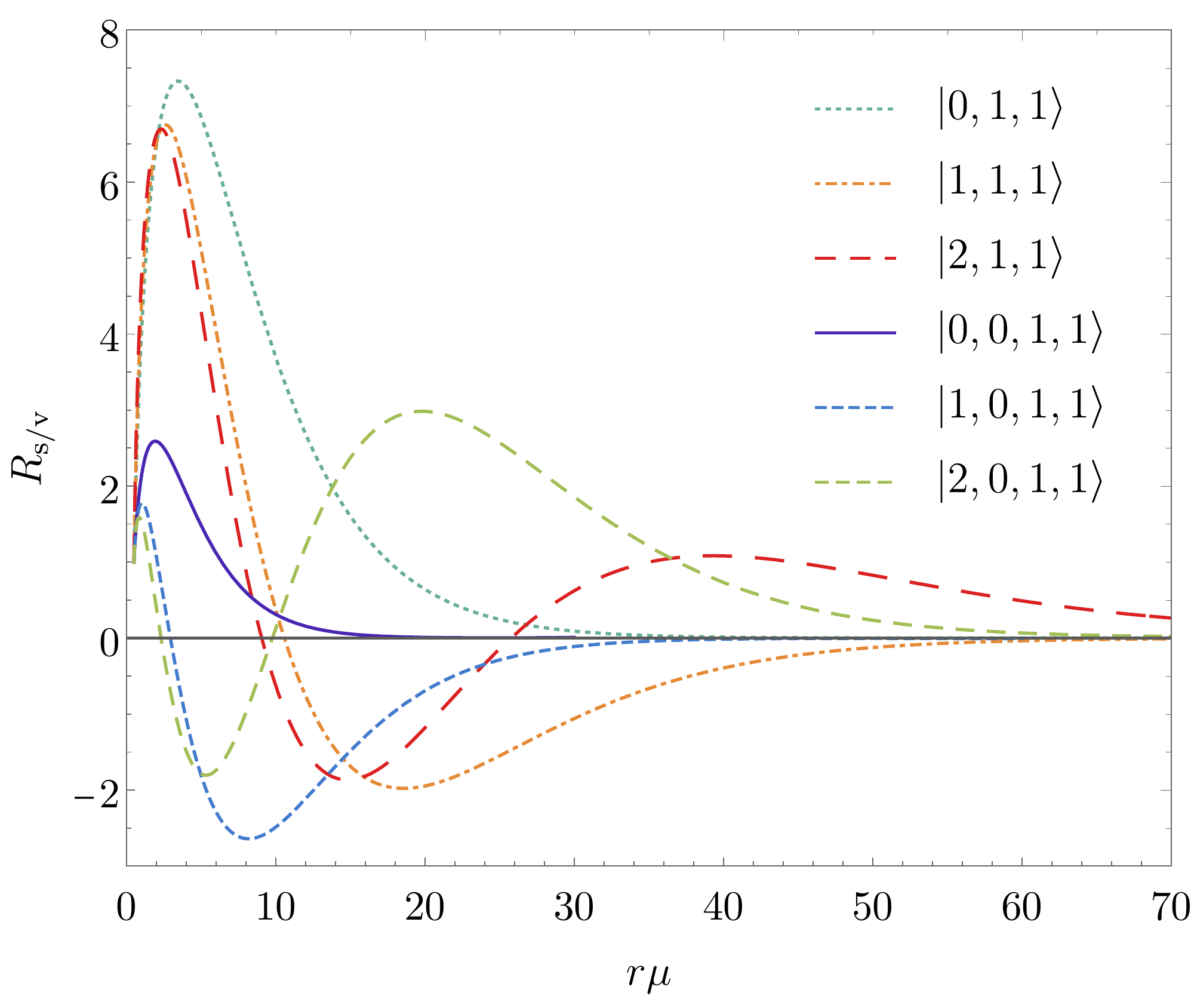}
  \caption{Radial profiles of the states marked in \autoref{fig:fig3} (bottom panel), characterized by $\mu r_\text{+}=0.5$. The radial functions are normalised so that $R(r_+)=1$.}
  \label{fig:fig4}
\end{figure}

A similar rationale holds true when fixing $\{\ell,j,m_j\}$ and varying $n$, as shown in \autoref{fig:fig3} for the states $\ket{n,1,1}$ and $\ket{n,0,1,1}$, $n=0,1,2$. The node number $n$ plays a role somehow akin to that played by the orbital angular momentum $\ell$. Large--$n$ states require larger minimum horizon angular velocities for stationary equilibrium. Vector states are still less energetic than their scalar cousins. The radial profile of the states marked with bullets in \autoref{fig:fig3} is depicted in \autoref{fig:fig4}. These states exist for Kerr black holes with $r_+\mu=0.5$ and have $n+1$ extrema. However, while the extrema of scalar states decrease towards spatial infinity, those of vector states increase. Vector states thus have wider spatial distributions.

\begin{figure}[h]
  \centering
  \includegraphics[width=.48\linewidth]{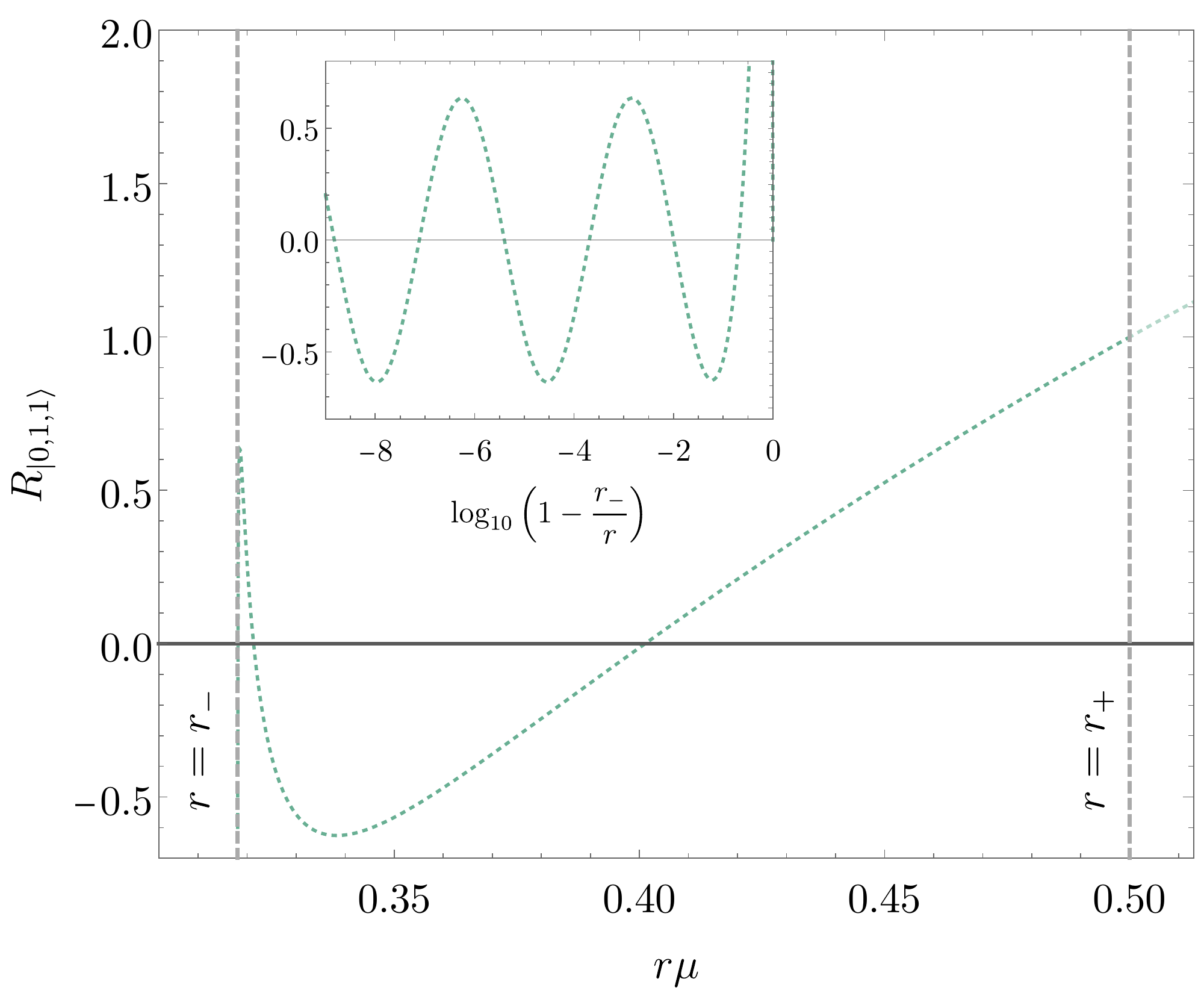}
  \quad
  \includegraphics[width=.48\linewidth]{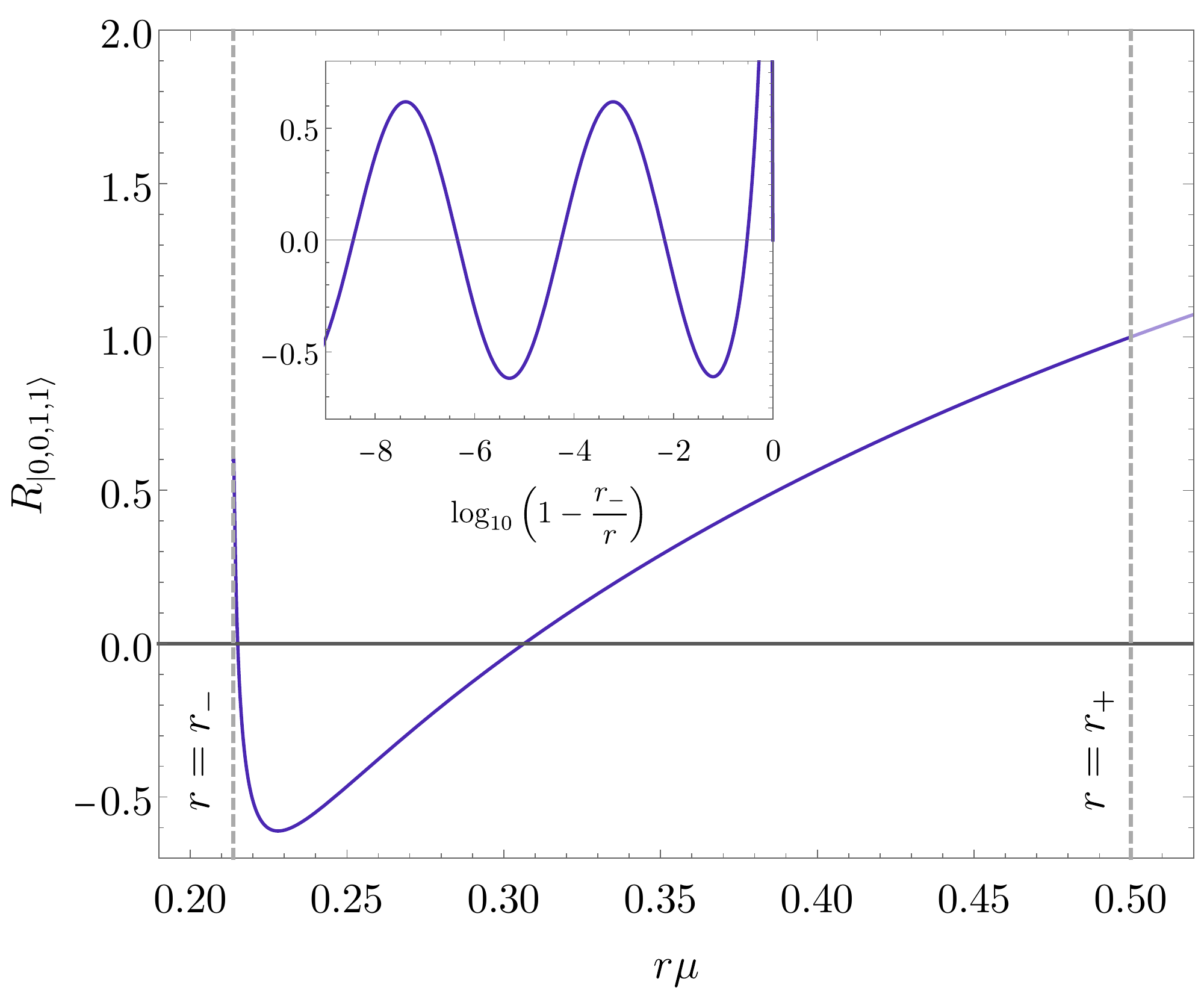}  
  \caption{Radial profiles of the states $\ket{0,1,1}$ (left panel) and $\ket{0,0,1,1}$ (right panel) marked in \autoref{fig:fig3} inside the Kerr black hole, characterized by $r_\text{+}\mu=0.5$. The radial functions are normalised so that $R(r_+)=1$.}
  \label{fig:fig5}
\end{figure}

The numerical solutions found using the direct--integration shooting method can be integrated from $r=r_+(1-\delta_+)$ to $r=r_-(1+\delta_-)$, with $\delta_-\ll1$. Since the appropriate boundary conditions at both the outer horizon and spatial infinity are already imposed, there is no freedom left to set the desired behavior at the inner horizon. The latter rotates with an angular velocity different from $\Omega_\hor$ and therefore synchronisation is not possible there. The radial profiles of the states $\ket{0,1,1}$ and $\ket{0,0,1,1}$ marked in \autoref{fig:fig3} is shown in \autoref{fig:fig5} in the black hole's interior. They exhibit oscillatory character close to $r=r_-$. This suggests that Kerr black holes with synchronised hair do not possess a smooth Cauchy horizon, but rather a curvature singularity at $r=r_-$~\cite{Brihaye:2016vkv}.

\subsection{Kerr--Newman black holes}

\begin{figure}[h]
  \centering
  \includegraphics[width=.28\linewidth]{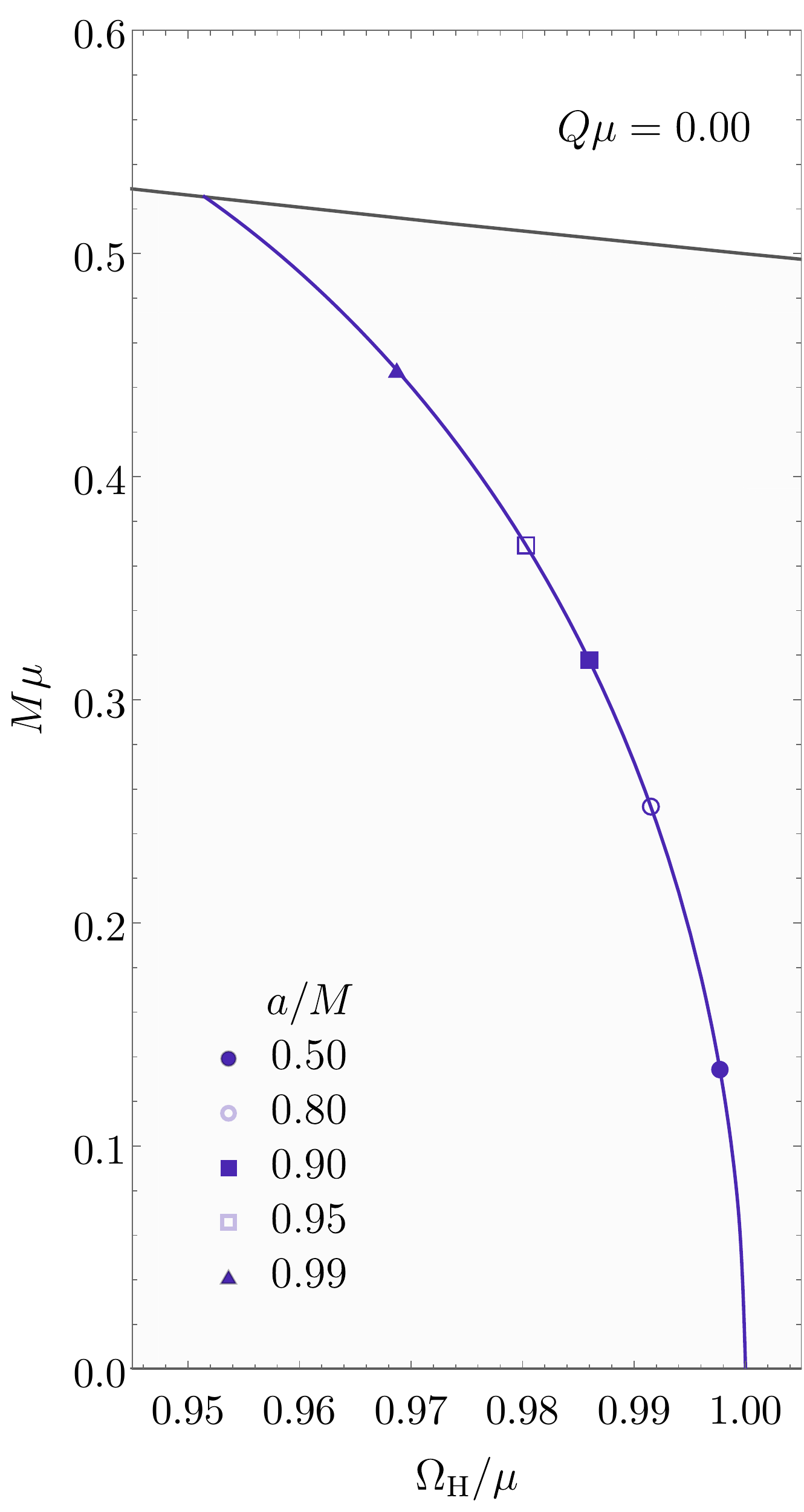}
  \hspace{-2.3em}
  \includegraphics[width=.28\linewidth]{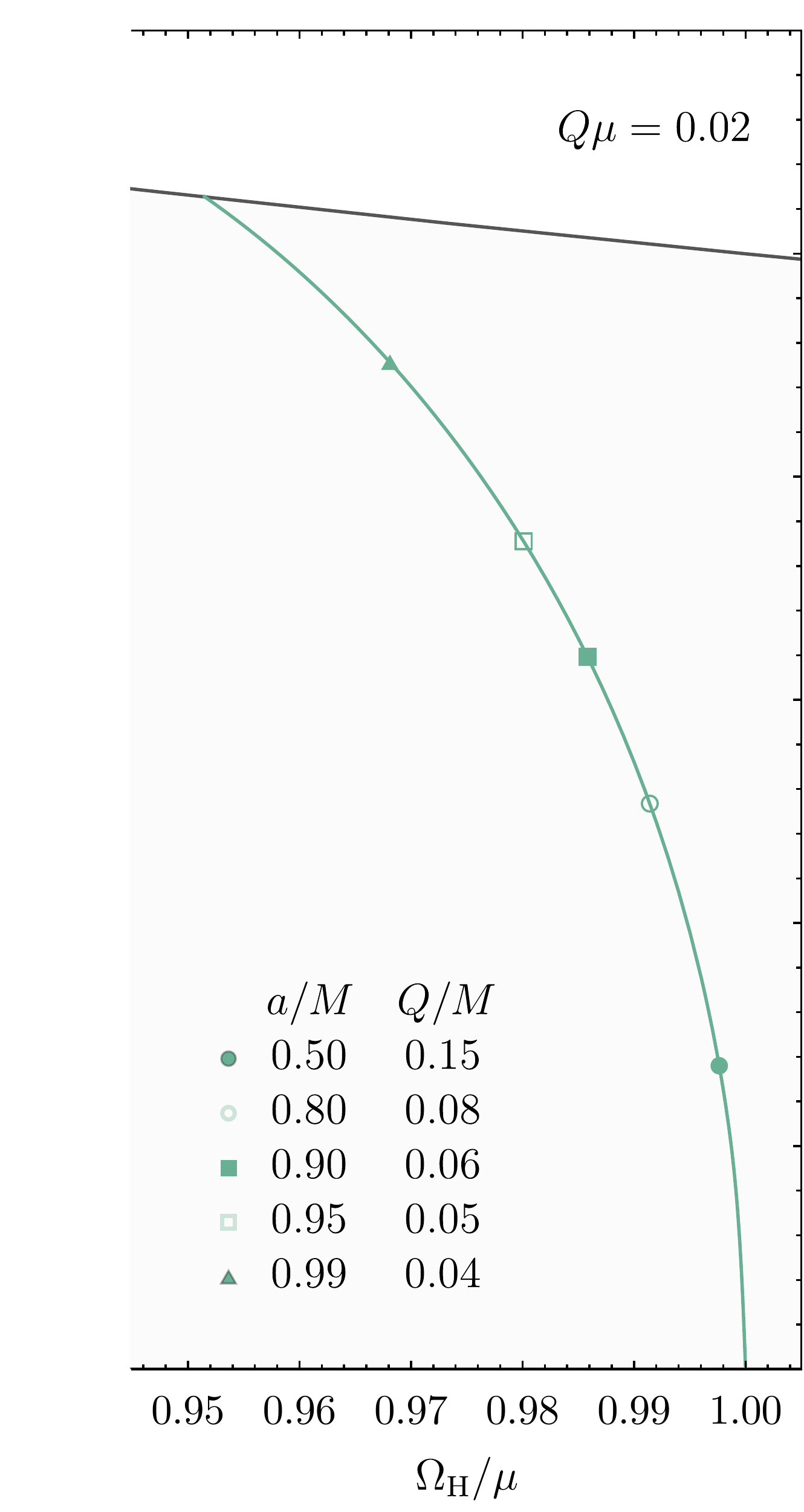}
  \hspace{-2.3em}
  \includegraphics[width=.28\linewidth]{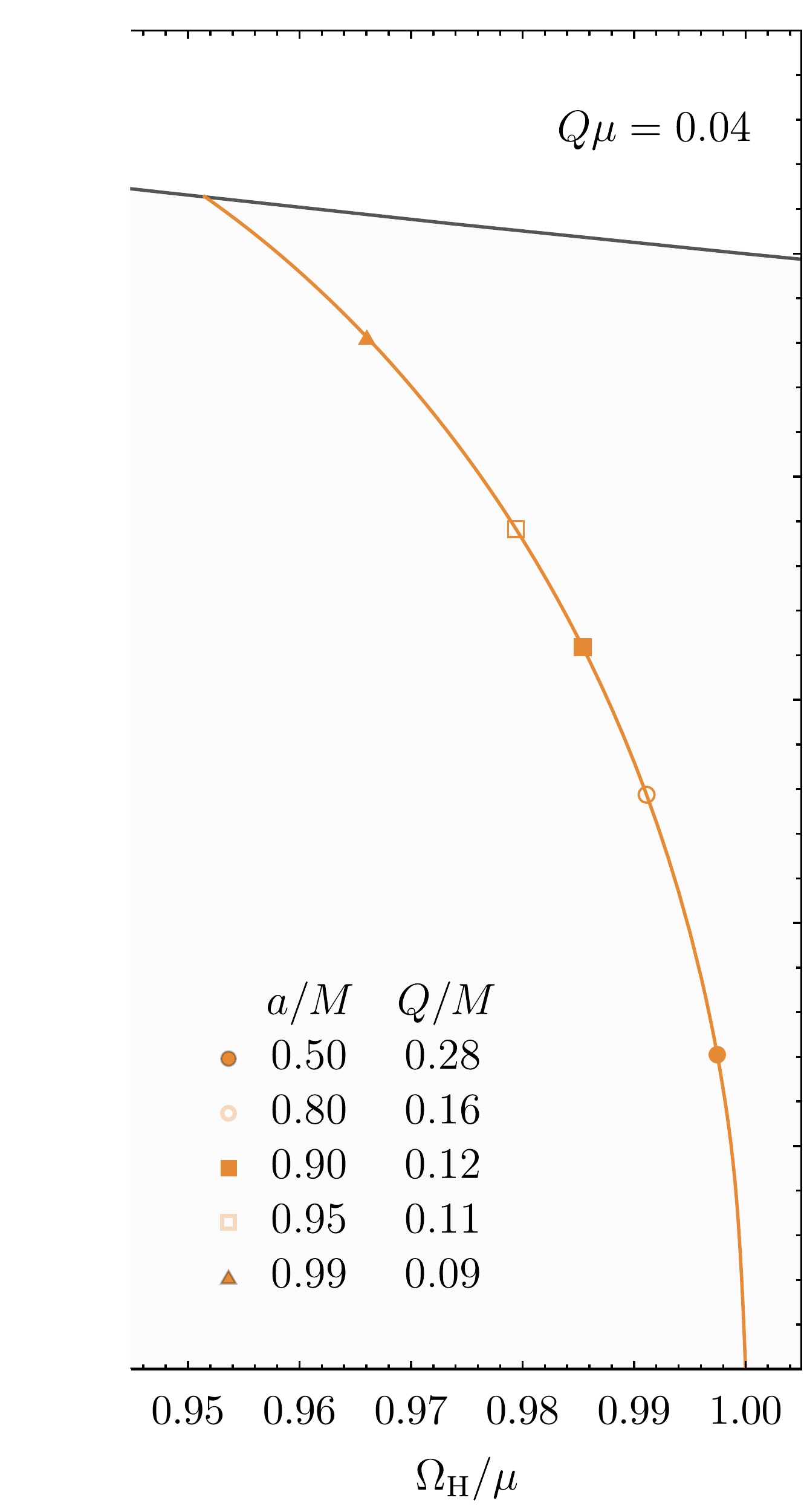}
  \hspace{-2.3em}
  \includegraphics[width=.28\linewidth]{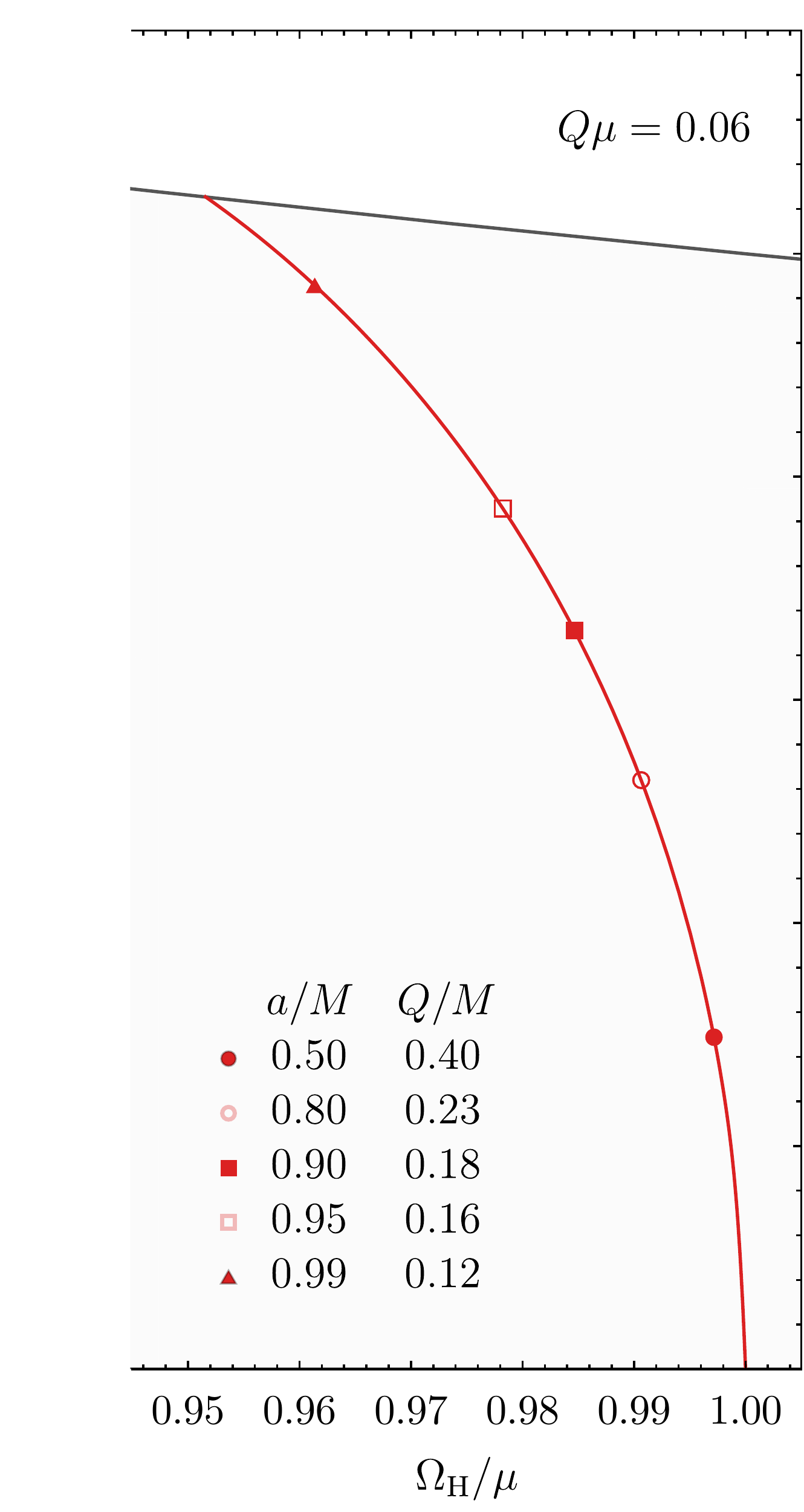}
  \caption{Existence lines of the scalar states $\ket{0,1,1}$ in the ($M\mu,\Omega_\text{H}/\mu$)--plane for Kerr--Newman black holes with different normalised charges $Q\mu$. The gray solid line refers to extremal ($a=M$) Kerr black holes. The specific electric charge is presented for black holes with different specific angular momenta.}
  \label{fig:fig6}
\end{figure}

The state of affairs does not change much when looking at synchronised states around Kerr--Newman black holes. The existence lines obtained when fixing the additional parameter $Q\mu$ coincide with those found for Kerr black holes ($Q\mu=0$) in the $(M\mu,\Omega_\hor/\mu)$--plane. However, each point on the line now represents a black hole with non-vanishing specific electric charge $Q/M$. Moving towards greater horizon angular velocities, the specific angular momentum $a/M$ decreases, whereas the specific electric charge $Q/M$ increases. The black holes close to the line $M\mu=0$ may be described as slowly-rotating extremal ($Q=M$) Reissner-Nordstr\"{o}m black holes. \autoref{fig:fig6} shows where Kerr--Newman black holes with $a/M\in\{0.50,0.80,0.90,0.95,0.99\}$ lie on the existence line of the scalar state $\ket{0,1,1}$ for different values of $Q\mu$. Similar trends are found for vector states. 

\newpage

\section{Conclusion\label{sec:5}}

This paper aimed at providing a comparative analysis of stationary scalar and vector clouds around Kerr and Kerr--Newman black holes. The key physical property of these bound states is a solidary rotation of the cloud with the black hole. These configurations are akin to the atomic orbitals of an electron in a hydrogen atom and can similarly be described in terms of $\{n,\ell,j,m_j\}$. This set of \textit{quantum numbers} label the existence lines of synchronised states in the parameter space of Kerr--Newman black holes and are continuously connected Kerr--Newman black holes with synchronised hair, solutions of Einstein--Maxwell theory minimally coupled to complex massive.

In general, vector bound states have greater energies than their scalar cousins and also occur for Kerr--Newman black holes in a wider domain of the normalised horizon angular velocity. The fundamental states match in both cases the most unstable quasi--bound state and are characterized by $j=m_j$, $n=0$ and the least possible value for the orbital angular momentum $\ell$. The latter two have similar impact on the cloud's energy for fixed $\{j,m_j\}$. Additionally, states with vanishing orbital angular momentum ($\ell=0$) are exclusive of vector bosons and are linked to a non--vanishing intrinsic angular momentum. 

The motivation behind a new glance at stationary clouds around Kerr--Newman black holes follows from the recent separation of the Proca equation in the  Kerr--NUT--(A)dS family of spacetimes. It would be of interest to apply the newfound ansatz to find synchronised states in other spacetimes and to construct stationary clouds in the time domain.

\section*{Acknowledgements}
The authors are especially grateful for the hospitality during their visit to Pará University, where part of this project was developed. 
This work is supported  by the Center for Astrophysics and Gravitation (CENTRA) and by the Center for Research and Development in Mathematics and Applications (CIDMA) through the Portuguese Foundation for Science and Technology (FCT -- Funda\c{c}\~ao para a Ci\^encia e a Tecnologia), references UIDB/04106/2020, UIDB/00099/2020 and UIDP/04106/2020. The authors acknowledge support  from the projects PTDC/FIS-OUT/28407/2017 and CERN/FIS-PAR/0027/2019 and from national funds (OE), through FCT, I.P., in the scope of the framework contract foreseen in the numbers 4, 5 and 6 of the article 23, of the Decree-Law 57/2016, of August 29,
changed by Law 57/2017, of July 19. This work has further been supported by  the  European  Union's  Horizon  2020  research  and  innovation  (RISE) programme H2020-MSCA-RISE-2017 Grant No.~FunFiCO-777740. The authors would like to acknowledge networking support by the COST Action CA16104.

\bibliographystyle{ws-ijmpd}
\bibliography{references}

\end{document}